\documentstyle{europhys}

\newif\ifboo \boofalse

\input psfig
\begin{document}

\euro{4x}{x}{x-x}{1999}
\Date{x xxxxxx 1999}
\shorttitle{BORIS~PODOBNIK {\it et al.}: SYSTEMS WITH CORRELATIONS IN 
THE VARIANCE}

\title{Systems with Correlations in the Variance: Generating Power-Law
Tails in Probability Distributions}

\author{Boris~Podobnik\inst{1,2}\footnote{Email:bp@phy.hr}, 
Plamen~Ch.~Ivanov\inst{1}, Youngki~Lee\inst{1},
Alessandro~Chessa\inst{1,3} and H.~Eugene~Stanley\inst{1}}

\institute{
        \inst{1} Center for Polymer Studies and Department of Physics,
                Boston University, Boston, MA 02215, USA\\
        \inst{2} Department of Physics, Faculty of Science, 
                University of Zagreb, Zagreb, Croatia\\ 
        \inst{3} Dipartimento di Fisica and Unit\'{a} {\it INFM}, 
                Universit\'{a} di Cagliari, 09124 Cagliari, Italy 
}

\rec{}{}

\pacs{
\Pacs{02}{50.Ey}{Stochastic processes}
\Pacs{05}{40.Fb}{Random walks and Levy flights}
\Pacs{05}{40.-a}{Fluctuation phenomena, random processes, noise, 
and Brownian motion}
}

\maketitle

\begin{abstract}
We study how the presence of correlations in physical variables
contributes to the form of probability distributions.  We investigate
a process with correlations in the variance generated by (i) a
Gaussian or (ii) a truncated L\'{e}vy distribution.  For both (i) and
(ii), we find that due to the correlations in the variance, the
process ``dynamically'' generates power-law tails in the
distributions, whose exponents can be controlled through the way the
correlations in the variance are introduced. For (ii), we find that
the process can extend a truncated distribution {\it beyond the
truncation cutoff}, which leads to a crossover between a L\'{e}vy
stable power law and the present ``dynamically-generated'' power
law. We show that the process can explain the crossover behavior
recently observed in the $S\&P500$ stock index.
\end{abstract}

\vspace{-1truecm}

Many natural phenomena are described by distributions with
scale-invariant behavior in the central part and power-law tails.  To
explain such a behavior the L\'{e}vy process \cite{L37} has been
employed in finance \cite{MB63}, fluid dynamics \cite{Swinney}, 
polymers \cite{Ott}, city growth
\cite{Makse}, geophysical \cite{Olami} and biological \cite{Viswan}
systems.  An intense activity has been developed in order to
understand the origin of these ubiquitous power-law distributions
\cite{Shlesinger-Klafter}.  The L\'{e}vy process, however, is
characterized by a distribution with infinite moments and in
applications this might be a problem, e.g. analysis of
autocorrelations in time series requires a finite second moment.  To
address this problem, probability distributions of the L\'{e}vy type
with both abrupt \cite {MS94} and exponential cutoffs \cite{K95} have
been proposed. A second problem is that, the L\'{e}vy process has been
introduced for independent and identically distributed stochastic
variables, while for some systems there is a clear evidence of correlations 
in the variance (e.g. for  many important market indices \cite {G77}).  
Moreover, a crossover between a L\'{e}vy stable power law and a power law 
with an exponent out of
the L\'{e}vy regime, was recently found in the analysis of price
changes \cite{Lux96}.  We investigate how a stochastic process with no
correlations in the variables but rather in their variance can be
introduced to account for the empirical observations of a L\'{e}vy
stable form of the probability distribution in the central part and a
crossover to a power-law behavior different than the L\'{e}vy in the
far tails.

In finance, a stochastic process with autoregressive conditional
heteroskedasticity (ARCH) \cite {Eng82} is often used to
explain systems characterized by correlations in the variance.  The
ARCH process is a discrete time process $x_t$ whose variance
$\sigma_t^2$ depends conditionally on the past values of $x_t$.  
The ARCH process is specified by the form of the
probability density function (PDF) $P(w_t)$ of the process $w_t$ 
which generates the random variable $x_t$.

Here we ask to what extent the presence of correlations in the
variance $\sigma_t^2$ contributes to the form of the probability
distribution $P(z_n)$, where $z_n \equiv \sum_{i=1}^{i=n} x_{t+i-1}$ and 
$x_t$ is the ARCH process
(in finance $z_n$ is called temporal aggregation of
the ARCH process $x_t$).  
We perform numerical simulations and investigate the form of $P(z_n)$,
when the PDF $P(w_t)$ of the process $w_t$ which generates $x_t$ 
is either of two cases, 
(i) a Gaussian, or
(ii) a truncated L\'{e}vy distribution.  For case (i) and $n=1$
(i.e. $z_1 \equiv x_t$), the
ARCH process generates $P(x_t)$ which, to a good approximation,
can be fit at the tails by a single power law \cite{SornetteD}.  
For case (ii) and $n=1$, we
show that the interplay between the L\'{e}vy form of the distribution
in the central part and the dynamics of the
ARCH process can give rise to a crossover between two
power-law regimes in the tails of $P(x_t)$.  For both (i) and
(ii), at large $n$, we find clear convergence of $P(z_n)$ to a Gaussian
\cite {footnote1}.

\begin{figure}
\centerline{\psfig{figure=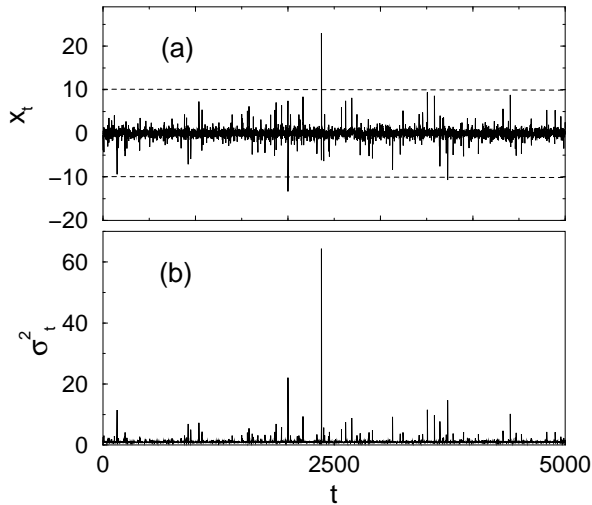,height=2.5truein,angle=-90}}
\vspace*{.3cm}
\caption{ 
The dependence on time of (a) a realization of an ARCH process $x_t$
of eq.~(\ref{uv}) and (b) the variance $\sigma_t^2$ of
eq.~(\ref{arch}).  The ARCH process $x_t$ is defined by $a = 0.88$ and
$b = 0.12$. For these $a$ and $b$ values, $\sigma^2_{x}=1$
(eq.~(\ref{var})). $P(w_t)$ is the TL PDF of
eq.~(\ref {TL}) with $\alpha=1.5$, $\gamma=0.27$, and the cutoff
length $\ell =10$ (horizontal dashed line).
Large values of $x_t$ are followed by large values of $\sigma_t^2$.  
Due to the dynamical features of the ARCH process,
values of $x_t$ can exceed 
the truncation cutoff $\ell$ of the TL distribution, and the 
regime beyond the cutoff length becomes populated [see fig.~3].}
%
\label{fig.1}
\end{figure}

Suppose $x_t$ is generated by an independent and identically distributed 
(i.i.d.) stochastic process $w_t$, 
drawn from a PDF $P(w_t)$ with zero mean $(<w_t>=0)$ and unit 
variance $(<w_t^2>=1)$,
\begin{equation}
x_t = \sigma_t  w_t.
\label{uv}
\end{equation}
Then $x_t$ follows an ARCH process if the variance $\sigma_t^2$
evolves in time as
\begin{equation}
\sigma_t^2 = a + b ~ x^2_{t-1}, 
\label {arch}
\end{equation}
where $a$ and $b$ are two non-negative constants \cite
{Eng82}.  For $b=0$, the ARCH process $x_t$ 
reduces to the i.i.d. stochastic process $w_t$ --- 
no correlations in all moments.  
For $b \neq 0$, the stochastic process
$x_t$ is uncorrelated --- $<x_t x_{t'}> \sim \delta_{tt'}$ ---
but has correlations in the variance $\sigma_t^2$. Indeed, the ARCH process 
is characterized by exponentially-decaying correlations in the variance
$<\sigma_t^2 \sigma_{t'}^2> \sim \exp\{-(t'-t)/\tau\}$ with decay constant
$\tau = |log(b)|^{-1}$. 

In fig.~1 we see that $\sigma_t^2$ is large when $x^2_{t-1}$ is large.  The
ARCH process 
is specified by the functional form of the 
PDF $P(w_t)$ which controls the stochastic variable
$w_t$  and also by the value of the parameter $b$ 
(eqs.~\ref {uv} and ~\ref {arch}). 
In the ARCH process $P(w_t)$ is 
traditionally the Gaussian,  but other choices are possible \cite{Boll87}.
Furthermore, the correlations in the variance $\sigma_t^2$
can be controlled by changing the parameter $b$ --- 
larger values of $b$ relate to stronger correlations.

\begin{figure}
\centerline{\psfig{figure=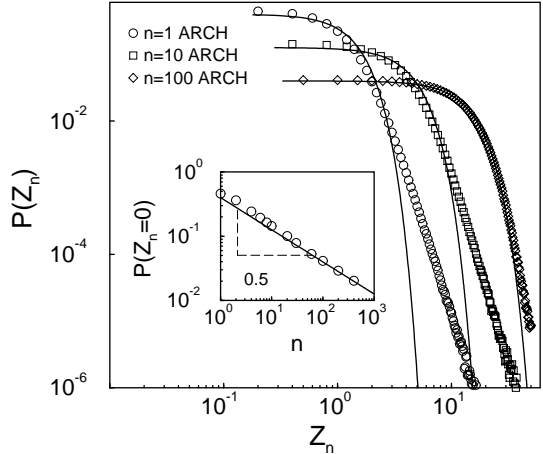,height=2.5truein,angle=-90}}
\caption{
Log-log plot of $P(z_n)$ for the temporal aggregation $z_n$ of 
the ARCH process $x_t$
calculated for different time scales $n=$1$(\circ)$, 
10$(\Box)$, and 100$(\diamond)$.  The ARCH process $x_t$ with 
$a=0.49$ and $b=0.51$ has the expected variance $\sigma^2_{x}=1$ 
(eq.~\ref {var}) and kurtosis $\kappa_{x}=10$ (eq.~\ref {kurtarch}). 
$w_t$ in eq.~(\ref {uv}) is chosen from a Gaussian distribution 
($\kappa_w=3$).  Due to the correlations in the variance $\sigma_t^2$,
the temporal aggregation of the ARCH process generates power-law
tails in $P(z_n)$ for small time scales $n$.
By solid lines, for the 
same time scales $(n=1, 10$, and $100)$, we denote the Gaussian 
process with the same variance as the ARCH process.  In the inset, we 
show a log-log plot of the probability of return to the origin $P(z_n=0)$
for the same ARCH $(\circ)$ and for the Gaussian process (solid line); 
the line has slope 0.5 --- the value predicted for the Gaussian --- 
which suggests that $P(z_n)$ gradually tends 
towards the Gaussian with increasing $n$.}
\label{fig.2}
\end{figure}

The expected variance of $x_t$ is a constant: 
$\sigma_{x}^2 \equiv <x_t^2>=<\sigma_t^2>$ \cite {Eng82}. From this
relation and eqs.~(\ref{uv}) and (\ref {arch}) follows
\begin{equation}
\sigma_{x}^2 = \frac{a} {1-b}
\label {var}
\end{equation}
and
\begin{equation}
\kappa_{x} \equiv \frac{<x^4>}{(<x^2>)^2} 
= \kappa_w + \frac {\kappa_w (\kappa_w-1) b^2}{1-\kappa_w b^2},
\label {kurtarch}
\end{equation}
where $\kappa_w$ is the kurtosis of $P(w_t)$, a common measure of 
the degree of ``fatness'' of the distribution related to the 
fourth moment \cite{G77,Eng82}.
From eq.~(\ref {var}), $\sigma_x^2$
is finite if $b < 1$, while finiteness of kurtosis $\kappa_x$ (and
the fourth moment of $x_t$) requires $\kappa_w b^2 < 1$.  No matter what
functional form for $P(w_t)$ we choose, the
ARCH process generates PDFs with slower 
decaying tails compared to $P(w_t)$ --- $\kappa_{x} > \kappa_w$ from
eq.~(\ref{kurtarch}). Such a slow decay in the tails of $P(z_n)$ for the 
ARCH process may come both
from the choice of the $P(w_t)$ and 
the presence of correlations in the variance, 
i.e. $\kappa_w$ and $b$ respectively.

First, we consider the temporal aggregation $z_n$ of an ARCH process 
for a particular choice of $P(w_t)$ given
by a Gaussian distribution (see fig.~2).  Due to the correlations in the
variance (eq.~(\ref{arch})), the ARCH process, for small $n$, generates
tails which can be approximated by a power law. 
Given that the parameter $b$ controls the correlations in the variance
$\sigma_t^2$, we find that the slope of this power law is directly linked
to these correlations --- the stronger the correlations in $\sigma_t^2$,
the smaller the slope of the power-law tail. 
For large $n$, the PDF of the temporal aggregation $z_n$ of the ARCH 
process resembles the form of the Gaussian distribution, 
i.e. $P(z_n)$ tends to the Gaussian, with a variance of $n \sigma_x^2$.

\begin{figure}
\centerline{\psfig{figure=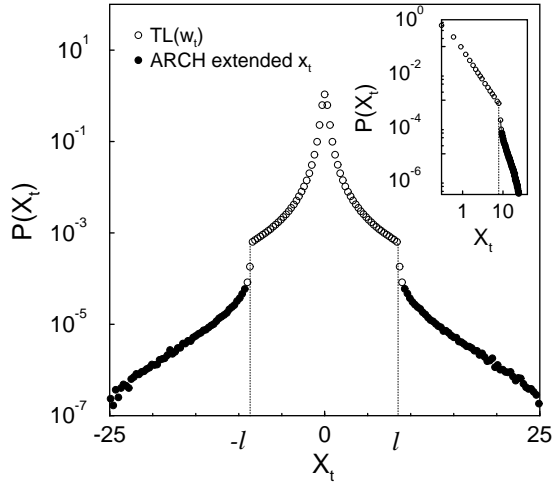,height=2.5truein,angle=-90}}
\vspace*{.3cm}
\caption{ 
Linear-log plot of the PDFs of $(\circ)$ the truncated L{\'e}vy
process $w_t$ without correlations and $(\bullet)$ the ARCH process
$x_t ~(n=1)$, where $w_t$ of eq.~(\ref {uv}) has the TL distribution
of eq.~(\ref {TL}) with $\alpha=1.2$, $\gamma=0.21$, and cutoff length
$\ell =9$ (vertical dotted lines), implying $\kappa_w=24.2$.
Parameters of the ARCH process are $a = 0.9$ and $b = 0.1$ giving
$\kappa_{x}=31.6$ (eq.~\ref {kurtarch}).  While the central part
$(\circ)$ of the PDF remains practically the same for both $w_t$ and
$x_t$ processes, the ARCH process $x_t$ extends the range of the TL to
$x_t > \ell$ due to the correlations in the variance $\sigma_t^2$
(eq.~\ref{arch}).
This ARCH-extended regime $(\bullet)$ is rarely populated and
therefore decays faster.  In the inset, we show log-log plot of the
PDF of the ARCH process $x_t$.  The process leads to extension of the
range of the original TL distribution, with a crossover between two
different power-law regimes: the original L{\'e}vy stable power law
and the dynamically-generated power law (appearing beyond the truncation
cutoff $\ell$).  The exponent of the power law in the ARCH-extended
regime is defined by correlations in the variance $\sigma_t^2$ through
the parameter $b$ (eq.~\ref{arch}).
}
\label{fig.3}
\end{figure}

Second, we analyze the ARCH process where $P(w_t)$ (eq.~(\ref {uv}))
is the truncated L\'{e}vy (TL) distribution \cite {MS94}, defined by
\begin{equation}
{\cal T}_{\alpha,\gamma,l} (w) \equiv 
\left \{ \begin{array}{c}
{\cal N}  {\cal L}_{\alpha,\gamma} (w)  ~~~~ |w|\le \ell
\nonumber\\
 0 ~~~~~~~~~~~~~~~ |w|> \ell
  \end{array}  \right \}.
\label{TL}
\end{equation}
Here, ${\cal L}_{\alpha,\gamma}(w)$ is the symmetrical L\'{e}vy stable
distribution \cite{L37,MB63}, $\alpha$ is the scale index $(0 <
\alpha < 2)$, $\gamma$ is the scale factor $(\gamma > 0)$, ${\cal N}$
is the normalizing constant and $\ell$ is the cutoff length.  Note,
that the TL distribution is characterized by power-law tails from zero
up to the cutoff length~$\ell$.  In performing numerical simulations
\cite {footnote}, we employ an algorithm of Ref. \cite {M94}.

In fig.~3, we show the PDF of an ARCH process $x_t$ ($n=1$) where
$P(w_t)$ is the TL distribution.  We find two regimes where the values
of $x_t$ are smaller and larger than the cutoff length $\ell$ of
$P(w_t)$.
The first regime ($x_t < \ell$) is characterized by the
L\'{e}vy power law due to the choice of the $P(w_t)$ (eq.~\ref{TL}).  
Due to the correlations in the variance $\sigma_t^2$
(eq.~\ref{arch}), the ARCH process extends the
range of the PDF to $x_t > \ell$, generating a new power law different
from the L\'{e}vy power law in the first regime.  
This second regime, where $x_t > \ell$, 
is rarely populated --- events in that regime occur only when
two large values of $x_t$ follow each other. Since
such events have small probability, the PDF of the ARCH process $x_t$
in that regime decreases faster compared with the first regime where
the fluctuations of $x_t$ are smaller than the cutoff length~$\ell$.

\begin{figure}
\centerline{\psfig{figure=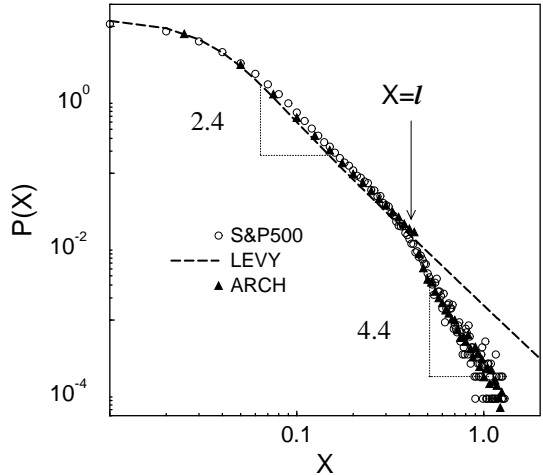,height=2.5truein,angle=-90}}
\vspace*{.2cm}
\caption{ 
PDF of the 1~min price changes for the $S\&P500$ stock index over the
12-year period Jan '84-Dec '95.  The central part of the empirical PDF
is well fit with the L{\'e}vy distribution ($\alpha=1.4$) up to the
cutoff length $\ell$ (arrow), followed by crossover to a second
``dynamically-generated by ARCH'' regime ($x_t > \ell$), which can be
approximated by a power law with slope $4.4$ ($\alpha=3.4$).  We show a
realization of the ARCH process, where $P(w_t)$ is the TL distribution
of eq.~(\ref {TL}) with $\alpha=1.4$, $\gamma=0.275$, length $\ell
=8$, and the process is characterized by $b=0.4$, where $a$ is chosen
to give the empirical standard deviation $\sigma=0.07$.  Note that the
bump at the truncation cutoff $\ell$ in the PDF of the ARCH process in
fig.~3 is more pronounced compared to the bump in the PDF of the
$S\&P500$ data and the ARCH process in fig.~4.  This is due to the fact
that the correlations in the variance $\sigma_t^2$ 
introduced for the particular realization of the
ARCH process in fig.3 are weaker ($b=0.1$) compared to the
correlations in the $S\&P500$ data ($b=0.4$).}
\label{fig.4}
\end{figure}

The existence of two regimes in the PDF, characterized by two
different power laws, is empirically observed in high-frequency data
on price changes \cite{Lux96}. In addition the central part of the
empirical PDF in financial data is well described by a 
L{\'e}vy distribution and exhibits the same scale-invariant
behavior \cite{MB63}.
Such behavior of the empirical PDF can be
mimicked by the ARCH process with a TL distribution, since the ARCH 
process generates power laws in the far tails while preserving the 
form of the PDF in the central region (figs.~2 and~3).
Empirical data
also show a bump in the PDF of the price changes calculated for a
delay of 1 min. A similar bump is observed in fig.~3 for the ARCH
process $x_t$ around the cutoff length $\ell$.  We find that this bump
is a signal for changing the power-law exponent in the probability
distribution.  In fig.~4 we show the PDF of the ARCH process in good
agreement with the PDF of 1~min price changes for the $S\&P$ 500
index. The central part ($x_t < \ell$) of the PDF characterized by
L{\'e}vy-stable power law with slope $2.4$ ($\alpha=1.4$) is followed by a
crossover to a second ``dynamically-generated'' by ARCH regime ($x_t >
\ell$) with a different behavior, approximately power law with
slope $4.4$ ($\alpha=3.4$).  
As in the case of the ARCH process with Gaussian PDF,
we can control the exponent of the power law in the second
dynamically-generated by ARCH regime by tuning $b$, while the
parameter $a$ is chosen to give the empirical standard deviation.

Next we study the asymptotic behavior (at large time scales $n$) of
the PDF of the temporal aggregation $z_n$, where $P(w_t)$ is the TL
distribution.  In fig.~5 we show $P(z_n=0)$ for the ARCH process.  For
small $n$, $P(z_n=0)$ approximately follows the L\'{e}vy distribution
${\cal L}_{\alpha, \gamma} (z_n = 0) = \Gamma(1/\alpha)/[\pi \alpha 
(n~\gamma)^{1/\alpha}]$
with the same $\alpha$ and $\gamma$ as $P(w_t)$. 
For $n > 30$, $P(z_n = 0)$ tends
to a Gaussian distribution $G(z_n=0) = 1/[\sqrt {2\pi} \sigma_{x} n^{1/2}]$
with variance $\sigma_{x}$ defined by eq.~(\ref {var}) and 
equal to the variance of $P(x_t)$ \cite {DN93}.
Thus, despite the presence of correlations in the variance $\sigma_{t}^2$,
the temporal aggregation $z_n$ of 
the ARCH process behaves like an i.i.d. process, characterized by a 
fast transition from L\'{e}vy to Gaussian process 
(``ultra-fast'' TL flight \cite{MS94}).
Since the correlations in the variance are 
of short range, for sufficiently large time scales $n$,
$P(z_n)$ approaches a Gaussian form, a behavior normally 
associated with uncorrelated stochastic variables (fig.~5).

\begin{figure}
\centerline{\psfig{figure=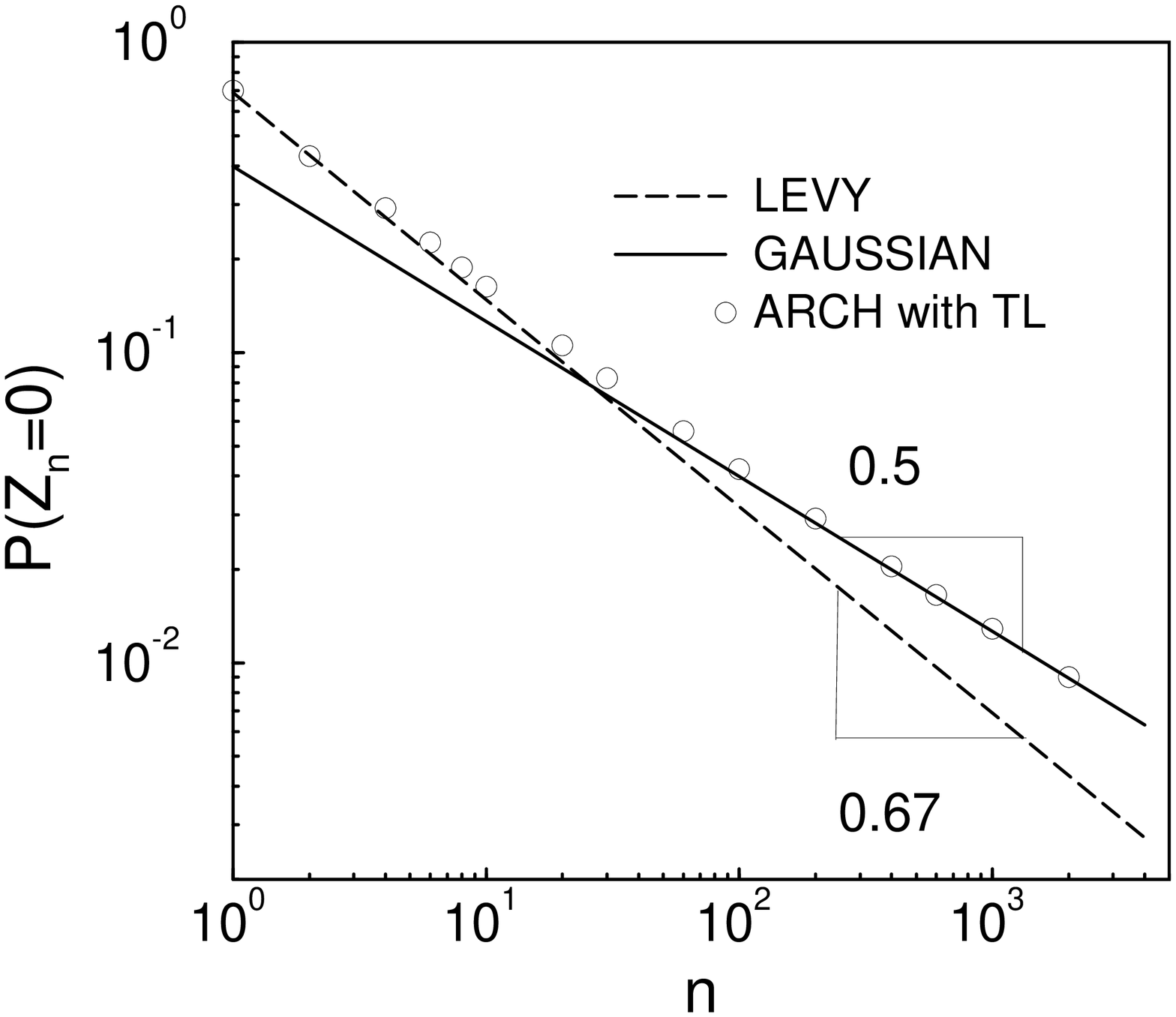,height=2.5truein,angle=-90}
\hspace*{-0.5cm}
\psfig{figure=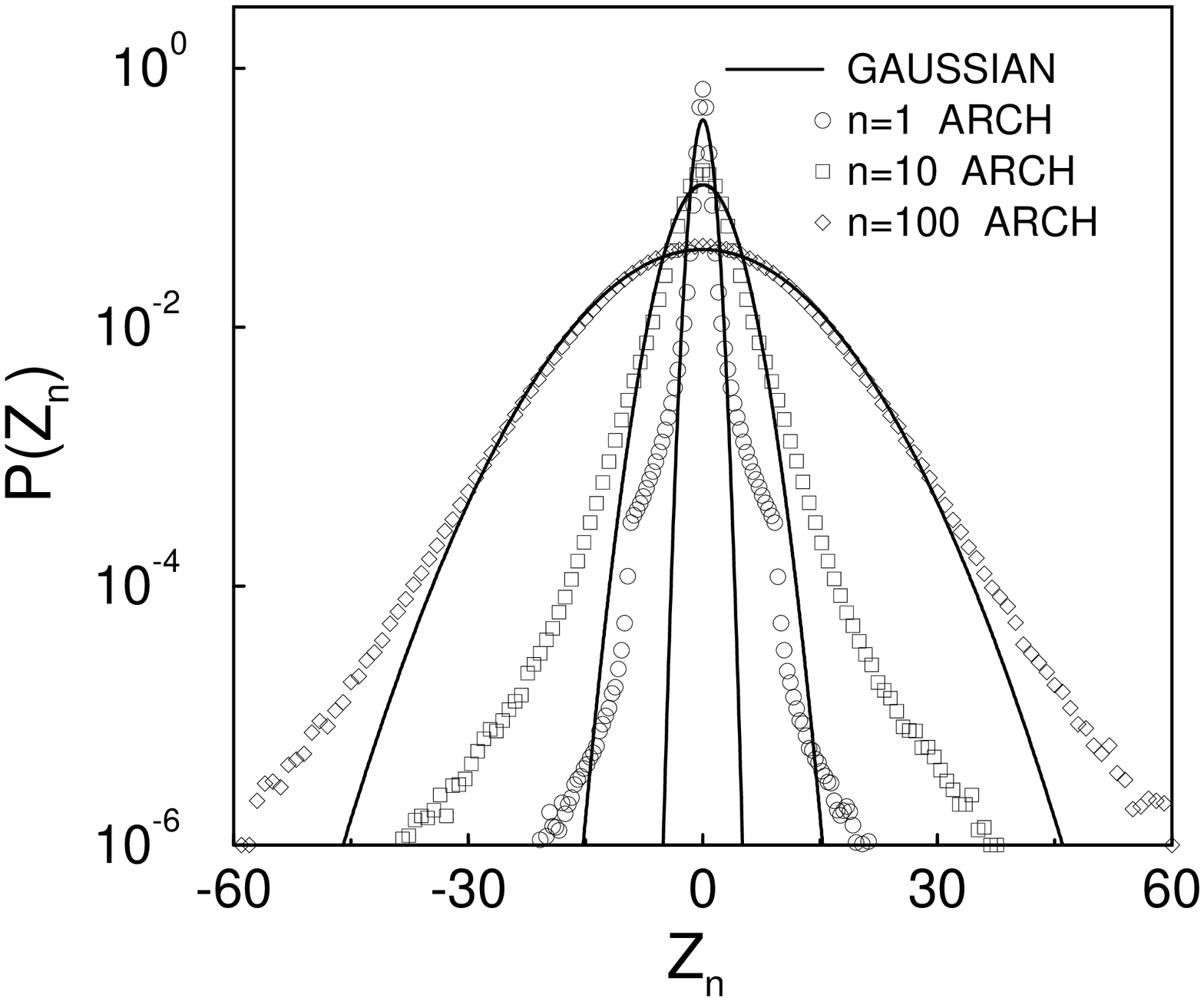,height=2.5truein,angle=-90}}
\caption{ 
(Left panel)
Log-log plot of the probability of return $P(z_n=0)$, where $z_n$ is
the temporal aggregation of the ARCH process
with $a = 0.88$, $b = 0.12$, $\sigma_{x}^2=1$
(eq.~\ref {var}) and kurtosis $\kappa_{x}= 31.4$ (eq.~\ref
{kurtarch}).  $P(w_t)$ is the truncated  L\'{e}vy distribution 
(eq.~\ref {TL}) (with $\sigma^2_w=1$ and $\kappa_w=21.9$ (eq.~\ref
{kurtarch})) obtained by setting $\alpha=1.5$, $\gamma=0.27$, and
$\ell =10$.  Also shown is the L\'{e}vy stable process  
with $\alpha=1.5$ (slope equals $1/\alpha$) and $\gamma=0.27$,
and the Gaussian process with variance $\sigma^2=1$
equal to the expected variance $\sigma_{x}^2$ of the ARCH
process.  For small values of $n$, $z_n$ follows the L\'{e}vy process
and then converges to the Gaussian.
(Right panel)
Linear-log plot of $P(z_n)$ calculated for different time scales $n$.
For the same time scales, we also show the Gaussian process 
to which $z_n$ gradually converges.
Parameters for both Gaussian and ARCH process are defined as for the 
left panel.}
\label{fig.5}
\end{figure}

In summary, we find that power-law tails in distributions can be
dynamically generated by introducing correlations in the variance of
stochastic variables, even when the initial distribution of these
variables is the Gaussian.  We also find that when the initial
distribution is a truncated L\'{e}vy distribution, the process of introducing
correlations in the variance can extend the range of the PDF beyond
the truncation cutoff.  This extension of the range of the original
probability distribution is characterized by a crossover between two
different power law regimes: the original L{\'e}vy stable power law
(within the limits of the truncation cutoff) and the
``dynamically-generated'' power law (beyond the truncation cutoff). This
behavior appears to explain the empirically-observed crossover in the PDF of
price changes for the $S\&P500$, and carries
information about the relevant parameters of the underlying stochastic
process. Our findings can help understand to what extent the presence of
correlations in physical variables contributes to the form of 
probability distributions, and what class of stochastic processes could
be responsible for the emergence of power-law behavior.

\centerline{***}

We thank Jan W. Kantelhardt and Kaushik Matia for helpful discussions,
and NIH/National Center for Research Resources (P41 RR13622) for
financial support.



\begin{thebibliography}{99}
\bibitem{footnote} 
To obtain a random variable with $\gamma \ne 1$, we note that the PDF
of eq.~(\ref {TL}) rescales under the transformations $ \tilde w
\equiv w /\lambda^{1/\alpha}$,~~ $\tilde {\cal T}_{\alpha, \tilde
\gamma,\tilde l }(\tilde w) \equiv {\cal T}_{ \alpha, \gamma, l }(w)
\lambda ^{1/\alpha}$, where $\tilde \gamma \equiv \gamma /\lambda $,
$\tilde l \equiv l/ \lambda ^{1/\alpha}$, and $\lambda$ is the scaling
parameter.  If any algorithm corresponding to random variable $w$ with
$\gamma = 1$ is provided, the scaling transformation allows us to
obtain any other variable $\tilde w$ with $\tilde \gamma \ne 1$.
\end{thebibliography}
\end{document}